\documentclass[nofootinbib,superscriptaddress,a4paper,twocolumn, floatfix,aps]{revtex4-2}
\usepackage{chemfig}
\usepackage[english]{babel}
\usepackage[utf8]{inputenc}
\usepackage{blindtext}
\usepackage{amsthm}
\usepackage{mathtools}
\usepackage{xcolor}
\usepackage{graphicx}
\usepackage[left=23mm,right=13mm,top=35mm,columnsep=15pt]{geometry} 
\usepackage[caption=false]{subfig}
\usepackage{algpseudocode}
\usepackage{bm}
\usepackage{adjustbox}
\usepackage{placeins}
\usepackage[T1]{fontenc}
\usepackage{lipsum}
\usepackage{float}
\usepackage{csquotes}
\usepackage[pdftex, pdftitle={Article}, pdfauthor={Author}, colorlinks, linkcolor={unia}, citecolor={unia}, urlcolor={unia}]{hyperref}
\definecolor{unia}{HTML}{3C2673}
\definecolor{fai}{HTML}{009b5d}
\definecolor{fai2}{HTML}{66c39e}
\definecolor{fai3}{HTML}{b3e1ce}
\definecolor{unia2}{HTML}{8c358b}
\definecolor{unia3}{HTML}{dbbfdd}
\definecolor{unia4}{HTML}{eddeee}

\hypersetup{
    colorlinks,
    linkcolor={unia},
    citecolor={unia},
    urlcolor={unia}
}
\usepackage{booktabs}
\usepackage{multirow}
\usepackage{listings}

\definecolor{codegreen}{rgb}{0,0.6,0}
\definecolor{codegray}{rgb}{0.5,0.5,0.5}
\definecolor{codepurple}{rgb}{0.58,0,0.82}
\definecolor{tqblue}{HTML}{08293d}
\definecolor{backcolour}{HTML}{fefdf5}

\lstdefinestyle{mystyle}{
    backgroundcolor=\color{backcolour},   
    commentstyle=\color{codegreen},
    keywordstyle=\color{magenta},
    numberstyle=\tiny\color{codegray},
    stringstyle=\color{codepurple},
    basicstyle=\ttfamily\footnotesize\color{tqblue},
    breakatwhitespace=false,         
    breaklines=true,
    postbreak=\mbox{\textcolor{magenta}{$\hookrightarrow$}\space},                 
    captionpos=b,                    
    keepspaces=true,                 
    numbers=left,                    
    numbersep=5pt,                  
    showspaces=false,                
    showstringspaces=false,
    showtabs=false,                  
    tabsize=2
}

\newcommand{\abs}[1]{\lvert #1 \rvert}

\newcommand{\ket}[1]{\lvert #1 \rangle}
\begin{document}

\title{Shallow Quantum Circuits for Deep Chemistry via Valence Bond Embeddings}
\author{Francisco Javier del Arco Santos}
\affiliation{{Institute for Computer Science, University of Augsburg, Germany }}

\author{Jakob~S.~Kottmann}
\email[E-mail:]{jakob.kottmann@uni-a.de}
\affiliation{{Institute for Computer Science, University of Augsburg, Germany }}
\affiliation{{Center for Advanced Analytics and Predictive Sciences, University of Augsburg, Germany }}

\date{\today}
\begin{abstract}
Quantum chemistry is one of the major potential applications in quantum computation. Currently there is a considerable focus on relatively small active spaces as a consequence of hardware noise and exponential bottlenecks in simulations. In the long run, there will be an increasing demand in reliable approximations for larger systems -- both, as initial states for projective algorithms like the quantum phase estimation or for the evaluation of dynamical properties.
While numerous approaches to select active spaces and extrapolate basis set accuracy exist, there is currently no consistent approach that results in a single quantum circuit for the total system.
In this work, we combine hybrid Fermionic-Bosonic encodings with the structured approach of Quantum Valence Bond Theory to directly construct quantum circuits for comparably large molecular systems. 
With this approach we are able to push simulability barrier of variational quantum eigensolvers towards chemically relevant systems and demonstrate circuit designs that outperform active space counterparts and achieve good approximations with respect to the exact solutions.
\end{abstract}

\maketitle
\begin{figure*}[t!]
    \centering
    \includegraphics[width=0.9\linewidth]{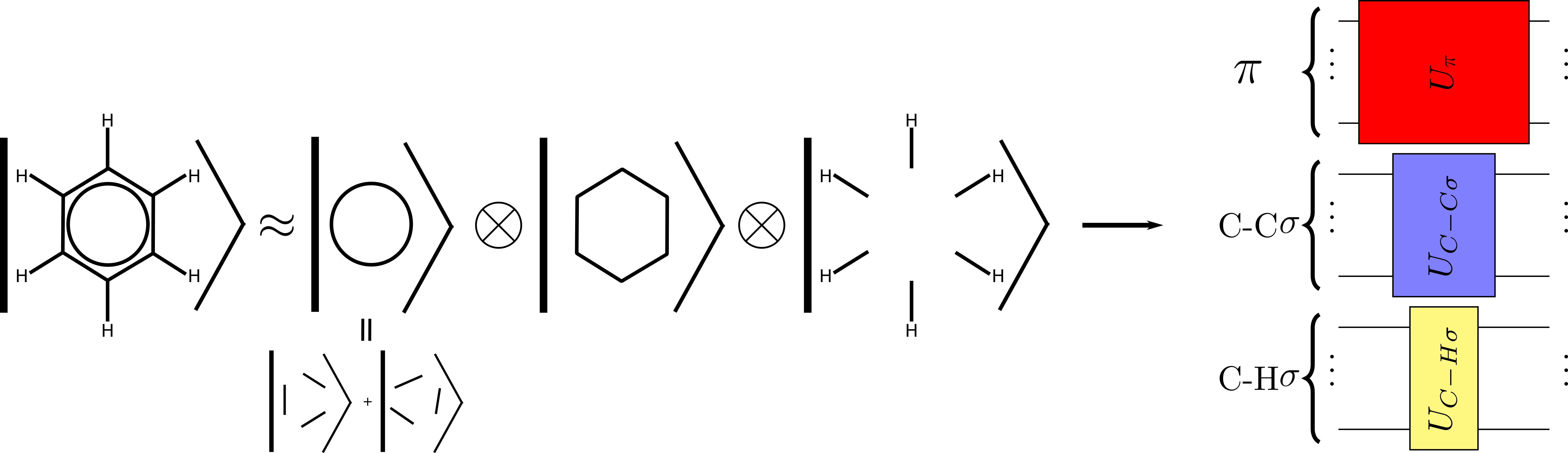}
    \caption{Illustration of the valence bond embedding on the example of a benzene molecule.}
    \label{fig:sep_circuit}
\end{figure*}
Recent years have witnessed growing interest in quantum computing as a potential paradigm shift for electronic structure calculations, promising to overcome the exponential scaling that limits classical methods when exact solutions are sought. While long term approaches centre around the quantum phase estimation~\cite{aspuru2005simulated} that projects an initial quantum state onto an eigenstate of the electronic system,~\cite{ollitrault2024Enhancing,fomichev2024Initial} direct approaches to prepare approximations of such states have heavily been investigated in the last decade. Most prominently are Variational Quantum Eigensolvers (VQEs) that leverage parametrized quantum circuits. However, despite initial~\cite{lee2018generalized} and recent~\cite{grimsley2019adaptive, burton2023Exact, burton2024Accurate, kottmann2022Optimized, ramoa2025reducing} progress in circuit design, practical applications of quantum computing in chemistry remain largely restricted to small active spaces and minimal basis sets, even in simulations. This limitations in principle arise from a combination of factors affecting both applied and simulated approaches. The former is dominated by hardware noise and limited qubit counts, while the latter is mostly affected by convergence problems; both depending on the  underlying circuit design. As a result, scalable approaches that are capable to generate good approximations for larger organic molecular instances could not be demonstrated so far.

As a consequence of these issues, most works resort to active-space treatments, where automated or semi-automated active space selection protocols,~\cite{stein2016automated,ding2023Quantum,tarocco2025AEGISS,shirazi2025Performance,bao2019Automatic} can often be adopted directly. For many applications (e.g. time evolution where weakly correlated parts can quickly become strongly correlated) it can however be efficient to incorporate correlations on the non-active part that are beyond a frozen doubly occupied description of the quantum state.

In this context, encoding strategies that reduce the effective size of the Hilbert space without compromising essential physics are particularly appealing. With this motivation in mind, we recently proposed hybrid fermionic-bosonic encodings~\cite{delarcosantos2025Hybrid} with the prospect to enable a compact but flexible representation of electronic states. Such approaches can significantly reduce circuit depth and the number of non-commuting Hamiltonian terms, both of which are critical resources. However, the success of these methods depends crucially on how the restricted subspace is defined and how this choice is reflected in the structure of the quantum circuit.

Here, we develop the methodology to overcome these limitations through Quantum Valence Bond (QVB) theory~\cite{kottmann2022molecular, kottmann2024Quantum} to guide both subspace selection and circuit construction. By exploiting the chemically intuitive structure provided by QVB, we establish a more systematic and interpretable connection between the chosen encoding and the underlying electronic structure.

As a result, we demonstrate that this combined strategy allows us to extend quantum simulations toward larger, chemically more relevant systems to bridge the gap between proof-of-concept demonstrations and future applications.

\section{Separable Circuit Designs}\label{sec:Sep_Circuit_Design}
Following prior works~\cite{kottmann2020reducing, kottmann2022Optimized} for qubit encodings, and related works in classical quantum chemistry~\cite{vaish2026Reducinga,lange2020Activea}, we will adopt a separable circuit design strategy to separate different part of the molecular instance.
Within this framework, the overall unitary $U_\mathcal{G}$ on $N$ qubits is decomposed into a tensor product of $M$ non-overlapping subcircuits
\[\mathcal{G} = \left\{a_k\right\}_{k=1}^{M},\;\; a_k\subset \left[1,2,\dots, N\right],\;\; a_k \cap a_k=\emptyset, \]
\begin{equation} \hat{U}_\mathcal{G} = \bigotimes_{a\in \mathcal{G}} \hat{U}_a, \end{equation}
so that the electronic wave function will consist of a tensor product of smaller quantum states 
\begin{equation}
\ket{\Psi}= \hat{U}_\mathcal{G} \ket{0} = \bigotimes_{a\in \mathcal{G}}\ket{\psi_a}.  
\end{equation}
Similar approaches have been employed on Refs.~\cite{otten2022Localized,wang2025Nonunitary,yoshikawa2022Quantum,zhang2022Variational}, where the molecular space is split based on geometrical reasons or quantum informational metrics, performing more accurate methods inside these fragments, followed by an inter-subsystem coupling section. In our approach, instead on looking for a full resolution for all subspaces, we chose the selection of subspaces and approximations applied to them based on the valence bond resonance structures of the molecule at hand. A schematic example of this approach is shown on Figure~\ref{fig:sep_circuit}, where in order to study a benzene ring, the system can be separated into: $\pi$ system, carbon-carbon $\sigma$ framework, and carbon-hydrogen $\sigma$ bonds, ordered by expected contribution. This would lead to three disentangled circuits of different complexities. \\\\
Note that, the expectation values from such a quantum state  are in general not decoupled, as the electronic Hamiltonian can be subdivided into
\begin{equation}\label{eq:Hsep}
\hat{H} = \sum_{k\in \mathcal{G}} \hat{H}_k + \hat{R}
\end{equation}
where $\hat{H}_{k} {\in \mathcal{G}}$ are the parts of the Hamiltonian restricted to a subsystem in $\mathcal{G}$ and the residual $\hat{R}$ consist of all terms in the Hamiltonian that are supported on more than a single subsystem. The residual can however be represented as
\begin{equation}
    \hat{R} =\sum_{i}\left( \bigotimes_{k\in G} \hat{h}^{(i)}_k \right)
\end{equation}
where type and number of terms $h_k^{(i)}$ in the depend on the chosen representation, the default for qubits being tensor products of Pauli matrices -- so called Paulistrings.
Within the applied techniques in this work, the coupling terms can be further restricted by taking symmetries and restrictions of the individual subgroups into account (see further sections). In particular, the particle conservation symmetry will result into only pairwise coupled interaction terms.\\

An expectation value with respect to a  separable quantum state is
\begin{align}
    E &=  \langle \Psi \rvert \hat{H} \lvert \Psi \rangle \label{eq:E}\\
    &= \sum_{a\in\mathcal{G}} \langle \psi_a \rvert \hat{H}_a\lvert \psi_a \rangle + \sum_i \prod_{a\in\mathcal{G}} \langle \psi_a\rvert \hat{h}_a^{(i)} \lvert \psi_a \nonumber\rangle
\end{align}

It should be noted that these approaches may provide poor approximations for many systems if the inter-subspace coupling is not treated with sufficient accuracy. In such cases, the results become highly sensitive to the chosen subsystem partitioning and to the initial distribution of electrons to the clusters in  $\mathcal{G}$. The impact of the first factor can be quantitatively assessed by comparing against the energy of the full, non-partitioned system, provided that the reference calculation is performed at an adequate level of theory. In contrast, the effect of an improper initial electron assignment is more difficult to verify. Here, valence bond theory will provide  an essential guideline. Based on the Lewis rules that assign two electrons to an edge in the resonance graph, the chemically sound assignment to the 3 clusters in the example above would be: 6 electrons to the benzene ring (plus the number of active electrons chosen in the metal), 12 electrons to the carbon-carbon, 12 electrons to the carbon-hydrogen framework.  
\subsection{Hybrid-Encodings}\label{subsec:HybridEncoding}
\begin{figure*}[t!]
    \centering
    \includegraphics[width=.95\linewidth]{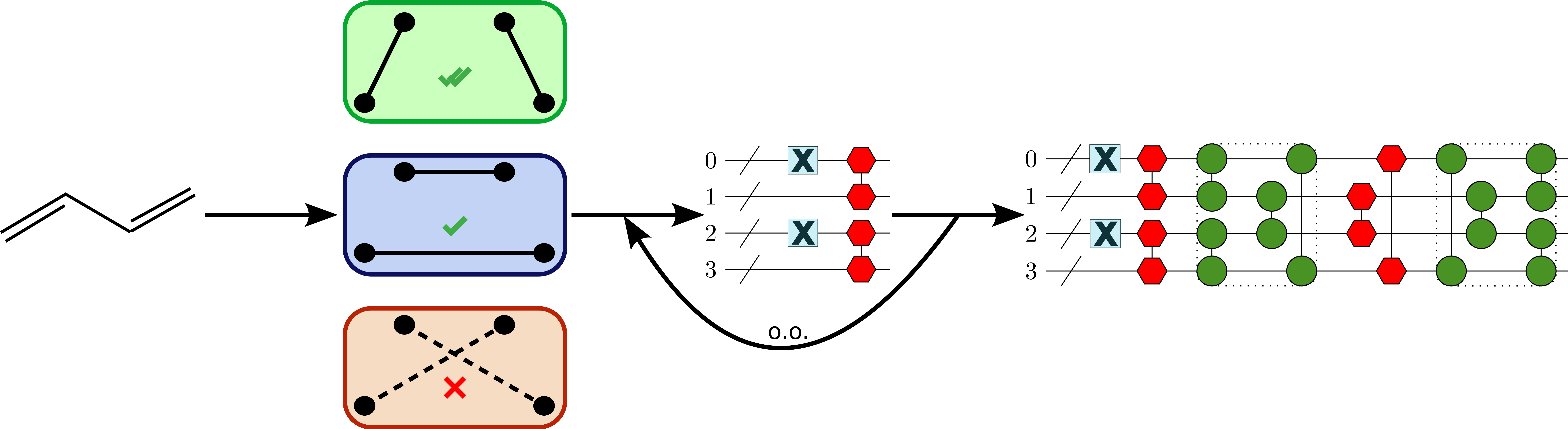}
    \caption{SPA+ circuit design employed in this work. The illustration is for the $\pi$-system of the butadiene molecule (see Fig.~\ref{fig:but_rotation} while the $\sigma$-framework is represented by a hardcore-bosonic circuit in separable pair approximation (SPA). The individual lines represent the 4 spatial $p_z$ orbitals that form the $\pi$ system (represented by 8 qubits). Hexagons represent spin-paired double-excitations while round gates represent orbital rotations. Orbital optimization (O.O.) is performed together with the $\sigma$-framework on the SPA circuit that reflects the primary resonance structure (in green) .}
    \label{fig:SPAplus_procedure}
\end{figure*}
Hybrid Fermionic-Bosonic encodings, introduced in a previous work~\cite{delarcosantos2025Hybrid}, partition the Fock space into two non-overlapping subspaces. The first retains a full Fermionic representation, while the second is treated as hard-core Bosonic (HCB) by forcing all occupancies to be spin-paired quasi-particles. 
Note that on this point, the electrons are still allowed to move freely between subspaces. Figuratively speaking: A HCB quasi-particle can leave the HCB subspace and split up into its spin-up and spin-down electrons and vice versa. In this formulation, the electronic Hamiltonian is split into three terms: one for each subspace and one interaction term
\begin{equation}\label{eq:Hybrid}
    \hat{H}= \hat{H}_{\mathcal{F}} + \hat{H}_{\mathcal{B}} + \hat{H}_I
\end{equation}
where $\hat{H}_{\mathcal{F}}$ denotes the Hamiltonian restricted to the Fermionic subspace, $ \hat{H}_{\mathcal{B}}$ the Bosonic subspace and $\hat{H}_I$ the interaction operator between both subspaces. Note that we have omitted the constant factor that, for example, arises from nuclear repulsion.\\
The individual operators are defined as
\begin{align}
\hat{H}_{\mathcal{F}}=& \sum_{i,j\in\mathcal{F}}h_{ij}\sum_{\sigma\in\{\uparrow,\downarrow\}}\hat{a}_{i\sigma}^\dagger \hat{a}_{j\sigma}\hspace{0.4cm}  \nonumber\\ &+\frac{1}{2}\sum_{i,j,k,l\in\mathcal{F}}g_{ij}^{kl}\sum_{\sigma\sigma'\in\{\uparrow,\downarrow\}}\hat{a}_{i\sigma}^\dagger \hat{a}_{j\sigma'}^\dagger \hat{a}_{k\sigma'}\hat{a}_{l\sigma},\label{eq:hfer}\\
    \hat{H}_B =& \sum_{i,j\in\mathcal{B}}(g_{ii}^{jj}+2h_{ij}\delta_{ij})\hat{b}_i\hat{b}^\dagger_j \hspace{0.4cm}\nonumber\\ &+\sum_{\substack{i\not=j\\i,j\in\mathcal{B}}}(2g_{ji}^{ij}-g_{ji}^{ji})\hat{b}^\dagger_i\hat{b}_i\hat{b}^\dagger_j\hat{b}_j,    \label{eq:hbos}\\
\hat{H}_I=&\sum_{\substack{i\in \mathcal{B}\\k,l\in \mathcal{F}}}(2g_{ik}^{li}+2g_{ki}^{il}-g_{ki}^{li}-g_{ik}^{il})\sum_{\substack{\sigma=\sigma'\\\sigma,\sigma'\in\{\uparrow,\downarrow\}}}\hat{N}_{i} \hat{a}_{k\sigma}^\dagger \hat{a}_{l\sigma'}\hspace{0.4cm} \nonumber\\ &+ \sum_{\substack{i\in \mathcal{B}\\k,l\in \mathcal{F}}}g^{kl}_{ii}\sum_{\substack{\sigma\not=\sigma'\\ \sigma,\sigma'\in\{\uparrow,\downarrow\}}}s_{\sigma\sigma'}\hat{b}^\dagger_i\hat{a}_{k\sigma}\hat{a}_{l\sigma'};\label{eq:hin}
\end{align}
with the usual second quantized  operators $\hat{a}_i$ ($\hat{a}^\dagger_i$) that annihilate (create) an electron in the $i$-th spin-orbital, the Bosonic operators $\hat{b}_i$ ($\hat{b}_i^\dagger$) that annihilate (create) a quasi-particle in the $i$-th spatial orbital, $\hat{N}_i$ the Bosonic number operator of the $i$-th spatial orbital ($\hat{N}_i = \hat{b}^\dagger_i\hat{b}_i$) and $s_{\sigma\sigma'}$ the factor
\begin{equation}
    \label{eq:sign}
    s_{\sigma_k,\sigma_l}= \sigma_k - \sigma_l, \quad \hspace{0.15cm}\sigma_k,\sigma_l\in\{\pm\frac{1}{2}\}.
\end{equation} 
 See also~\cite{henderson2015Pair} for an in-depth discussion of the Bosonic operators ~\cite{delarcosantos2025Hybrid} for more details on this hybrid encoding. \\
 The coefficients $h_{ij}$ and $g_{ij}^{kl}$ are the usual electronic integrals, here in the ``1221'' (or ``Google'') notation for the electron repulsion integrals
\begin{equation}
\label{eq:1eeint}
h_{ij}=\int\varphi_i^*(x)(-\frac{1}{2}\nabla^2-\sum_l\frac{Z_l}{r_l})\varphi_j(x)dx,
\end{equation}
\begin{equation}
\label{eq:2eeint}
g_{ij}^{kl}=\iint\frac{\varphi^*_i(x_1)\varphi^*_j(x_2)\varphi_k(x_2)\varphi_l(x_1)}{r_{12}}dx_1dx_2.
\end{equation}
Although the resulting Hamiltonian may appear more complex than the fully fermionic counterpart, the effective treatment of electrons in the “bosonic” orbitals relaxes the constraints imposed by fermionic antisymmetry. As a consequence, the number of non-commuting operator groups is reduced and the unitary excitation operators can be compiled to significantly shallower circuits.

\subsection{Separable Pair Approximations}\label{subsec:SPA}
For the Bosonic subspaces, adding another approximation on top has shown to behave in particularly well: The separable pair approximation, that not only forces electrons to form spin-paired quasi particles (monogamy principle), but also assumes that the pairs can be separated by a tensor product structure (strong monogamy principle). Technically this requires to assign each orbital to a specific electron pair which makes the method highly sensitive to the shape of the orbitals (see section on orbital optimization).\\\\ 
In the original SPA ansatz~\cite{kottmann2022Optimized}, the wave function is built as a tensor product of $N_e$-electron$/2$ pair functions:
\begin{equation}\label{eq:SPA}
    \ket{\Psi_{SPA}} = \bigotimes_{k=1}^{N_e/2}\ket{\psi_k}.
\end{equation}
Each $\ket{\psi_k}$ is a linear combination of tensor products of $\abs{S_k}$ one-electron products,
\begin{equation}
    \label{eq:pair_wfv}
    \ket{\psi_k}=\sum_{m,n\in SOs}c_{mn}^k\ket{\phi_m^k}\otimes\ket{\phi_n^k},
\end{equation}
each $\ket{\psi_k}$ represented its own set of orbitals $S_k=\{\phi_l^k,l=0,...,\abs{S_k}-1\}$. The summation is carried over Spin Orbitals (SOs), which are the Spatial Orbital (also denoted as Molecular Orbital, MO) with the extra spin coordinate.  At this point, one needs to store $\mathcal{O}(\abs{S_k}^2)$ coefficients; Within the HCB approximation, eq.~\ref{eq:pair_wfv} needs to fulfill the spin-pairing requirements:
\begin{equation}
    \label{eq:pair_wfv_hcb}
    \ket{\tilde{\psi}_k}=\sum_{m\in MOs}c_{m}^k\ket{\phi_{m\uparrow}^k}\otimes\ket{\phi_{m\downarrow}^k},
\end{equation}
lowering the memory requirement to $\mathcal{O}(\abs{S_k})$. In both cases, the wavefunction can be efficiently stored classically. A priori, there is no unique strategy to determine the orbital subspaces $S_k$; as an instance, one could define these sets as each canonical HF orbitals with its respective antibonding orbital, recovering part of the static correlation~\cite{henderson2014seniority}. However, one can take advantage of the concept of molecular graph (or Lewis structure) widely employed in Valence Bond theory (VB).~\cite{hiberty2026Mapping,ghasempouri2023Modulara} Similar to other VB methods on classical quantum chemistry, one starts by building the possible graphs following Rumers' rule.~\cite{karadakov2008Advances,simonetta1968ValenceBond,rumer1932spin} Then, the most relevant graph is chosen to build the ansatz, assigning $\abs{S_k}$ as the edges on the graph, further optimizing the orbitals to minimize this SPA's energy. Empirically we often observed, that there is little benefit to introduce interactions between the individual electron pairs as long as the HCB approximation is applied, the SPA was often observed to be a good approximation to UpCCD and its variants (see for example Ref.~\cite{kottmann2023Molecular} for studies comparing against UpCCD variants and adaptive circuit construction).\\\\

\subsection{Beyond Separable Pairs}
In the proposed Valence Bond Embeddings of this work the molecular system is split into a $\sigma$-framework approximated represnted by an SPA wave function in hardcore-bosonic encoding and one or more further resolved parts that are encoded with Fermionic modes.\\

Illustrating the concept on the example in Fig.~\ref{fig:sep_circuit},
we start by choosing appropriate clusters guided by valence bond resonance structures. Here we have $\pi$-system and the CC and CH $\sigma$ frameworks, leading to $\mathcal{G}=\left\{C_\pi, C_{\sigma_1}, C_{\sigma_2}\right\}$. The total wavefunction is then represented as
\begin{align}
    \ket{\Psi} = \bigotimes_{a\in \mathcal{G}} \hat{U}_a\ket{0}.
\end{align}
For each cluster we can make the choice to represented it in full Fermionic resolution or restrict it to hardcore Bosons, while for the latter we can, in addition also restrict the state $\ket{\psi_a} = U_a\ket{0}$ on the cluster to be represented by separable pairs. 
For the CH $\sigma$-framework this would for example lead to a tensor product of 6 2-electron states. As the CC framework might be more coupled to the $\pi$-system as the CH framework, we could also chose to combine the $\pi$-system with the CC framework, ending up with a total two clusters in $\mathcal{G}$.\\

While one can in principle resort to any type of circuit design within the fully resolved parts of the molecule (e.g. with $k$-UpCCGSD~\cite{lee2018generalized}, tUPS~\cite{burton2024Accurate} or  ADAPT~\cite{grimsley2019adaptive} or approaches based on Hamiltonian time-evolution~\cite{granet2024Hamiltonian} ), we will sketch in this section how this can be done with Valence Bond guided construction. Take for example the delocalized $\pi$-system of the Benzene ring in Fig~\ref{fig:sep_circuit}.\\\\
In order to significantly improve on the SPA model~\cite{kottmann2023Molecular} it is necessary to leave the HCB approximation as well as temporarily rotate the orbital frame through a corresponding circuit, while later rotating it back to the original frame. This encode-decode strategy can then be combined with the remaining resonance structures.~\cite{kottmann2023Molecular,kottmann2024Quantum} A schematic description of this procedure can be found in Fig.~\ref{fig:SPAplus_procedure}, where the graphs are encoded through orbital correlators, usually paired-double excitations, represented here by the red hexagon gates. In order to include further graphs, the basis must be changed to each graph's suitable basis through the application of a givens rotations layers (layers of paired single excitations)~\cite{google2020hartree}, which are represented here by the green circular gates. Then, further graphs are introduced by including orbital correlators, usually cheap paired double excitations, represented in Fig.~\ref{fig:SPAplus_procedure} by the red hexagons. Note that not all graphs will be included since the principles of Valence Bond Theory state that crossing graphs are a linear combination of other graphs, therefore they are redundant.~\cite{hiberty2026Mapping}
\\\\This approach shows how to sequentially increase the quality of our circuits by adding more resonance structures. One can, in principle, just brute-force it and add all possible graphs, or simply use the chemical intuition. Note however, that the convergence of the VQE procedure was observed to become more challenging the more resonance structures are added.\\\\
In order to automate the graph selection, Quantum Information Theory could be employed, extracting some metrics such as the Mutual Information or Orbital entanglement, similar to refs.~\cite{ding2023Quantum, zhang2020mutual} using a wave-function built on Hybrid Atomic Orbitals (HAOs) and classify all contributions by weight, ensuring every atom is only connected once. At this point, we will not use this approach, as it heavily relies on extracting the information from an already existing wave function.

\subsection{Orbital Optimization}\label{subsec:OO}
As discussed in the previous sections, electronic wavefunction in hardcore-bosonic approximations are sensitive to the choice of orbitals.~\cite{bytautas2011Seniority,limacher2014Influence} In order to find the best possible linear combination standard orbital-optimization methods can be used. While recent developments~\cite{limacher2026Attacking} specialized for hardcore-bosonic wave functions offer a interesting path to faster optimizations in general, they still can't guarantee convergence which will heavily depend on the chosen initial guess. In the following we will illustrate how we are constructing initial orbital guesses for wave functions generated through SPA circuits.\\\\
In this work, we will mostly restrict ourselves to Gaussian type minimal basis sets (STO-3G~\cite{hehre1969SelfConsistent}) as they offer a simple framework. In section~\ref{sssec:bsc} we will explore a possible path towards more accurate representations.\\ 

Once the basis is selected, the optimal Molecular Orbitals for the quantum circuit at hand have be determined as linear combinations of the basis orbitals. A widespread approach is to start from the canonical HF orbitals as initial guess, in the context of quantum circuits this strategy often fails due to the unstructured virtual (orbitals not occupied in the Hartree-Fock determinant) space. In this work, we will generalize the workflow presented in ref.~\cite{kottmann2022Optimized} that was based on pair-natural orbitals.\\

The algorithm can be roughly described as
\begin{enumerate}
    \item Build hybrid atomic orbitals (HAO) in sp, sp$^2$, or sp$^3$ hybridization and assign them to the vertices of the chemical graph
    \item Align the orbitals with the direction of the bonds (edges)
    \item Combine hybrid orbitals from connected vertices into bonding and anti-bonding types
    \item Use this orbitals as initial guess in an orbital optimizer
\end{enumerate}
For pure hydrogenic systems, this strategy can be implemented relatively straightforward (see for example ~\cite{bincoletto2026Transferable}). For heterogenic molecules the manual preparation of the initial guess is however tedious, which is why we resort to ``Chemistry Localized Property-optimized Orbitals'' 
(CLPO)~\cite{nikolaienko2019Localizeda} as the underlying design principles align well with the strategy above. These CLPO are obtained by decomposing a reference wave function (typically the Hartree-Fock state, but any other mono-reference state might be employed) into 1 and 2-center localized orbitals, imposing a double electron occupation of the resulting bonding orbital and zero for its anti-bonding counterpart. On this classical approach, the main graph is found by a maximum-weight matching “blossom algorithm” which maximizes the sum of all edges' weights, which are related to the 1-body Reduced Density Matrix. Given that the applicability of this approach relies on the initial guess wave function suitability and the pairing algorithm, both this approach and the orbital construction by HAO pairing will be employed on the following indistinguishably.
From now on, we will refer to these CLPOs optimized for SPA energy minimization as \textit{SPA orbitals} to differentiate them from other localization schemas.

\section{Valence-Bond-Embeddings}\label{sec:VB_embedings}
Summarizing the individual steps of the last section we arrive at a general procedure to construct separable circuits in hybrid encodings.
In the following we illustrate the valence bond embedding that is the central technique of this work. 

\begin{enumerate}
    \item Initialize the CLPO through the blossom algorithm referenced in the previous section. Construct a Lewis graph Based on the CLPO structure 
    \item[1.] (alternative) Select a Lewis graph for the molecule and manually configure the initial orbitals to resemble the graph structure. 
    \item Decompose the graph into subgraphs $C_k \in \mathcal{G}$ and select a representation (hardcore bosonic or fermionic). Assign all CLPO to vertices in the graph.
    \item Assemble SPA circuits for all hardcore bosonic parts
    \item Assemble VQE circuit for the Fermionic parts
    \item Compile the objective function~\eqref{eq:E} that represents the expectation value
    \item Execute the VQE optimization (either classically, on quantum hardware, or mixed)
    \item Optimize the molecular orbitals
\end{enumerate}

In this work we chose a multigraph schema, as illustrated in Fig.~\ref{fig:SPAplus_procedure} for the fermionic VQE parts in step 4. 
In the default construction illustrated above, all bosonic subspaces are classically simulable the underlying SPA circuit structure that restricts each bosonic subspace to a single quasi-particle. This limitation can be relaxed by introducing additional correlations, for instance by coupling multiple orbitals through an UpCCD layer, either in an all-to-all fashion or under a retained separability constraint. An advantage of such an extension is that it can be straightforwardly compiled into qubit-excitation gates. However, incorporating this additional flexibility would already compromise the classical simulability of this sector. It won't be done on this work unless otherwise stated. In the same way, some of the clusters can be encoded in a fermion-boson hybrid encoding as well, without imposing separability between them.

\section{Applications}
\begin{figure*}[t!]
    \centering
    \subfloat[\label{sfig:but_act}]{\includegraphics[width=0.49\linewidth]{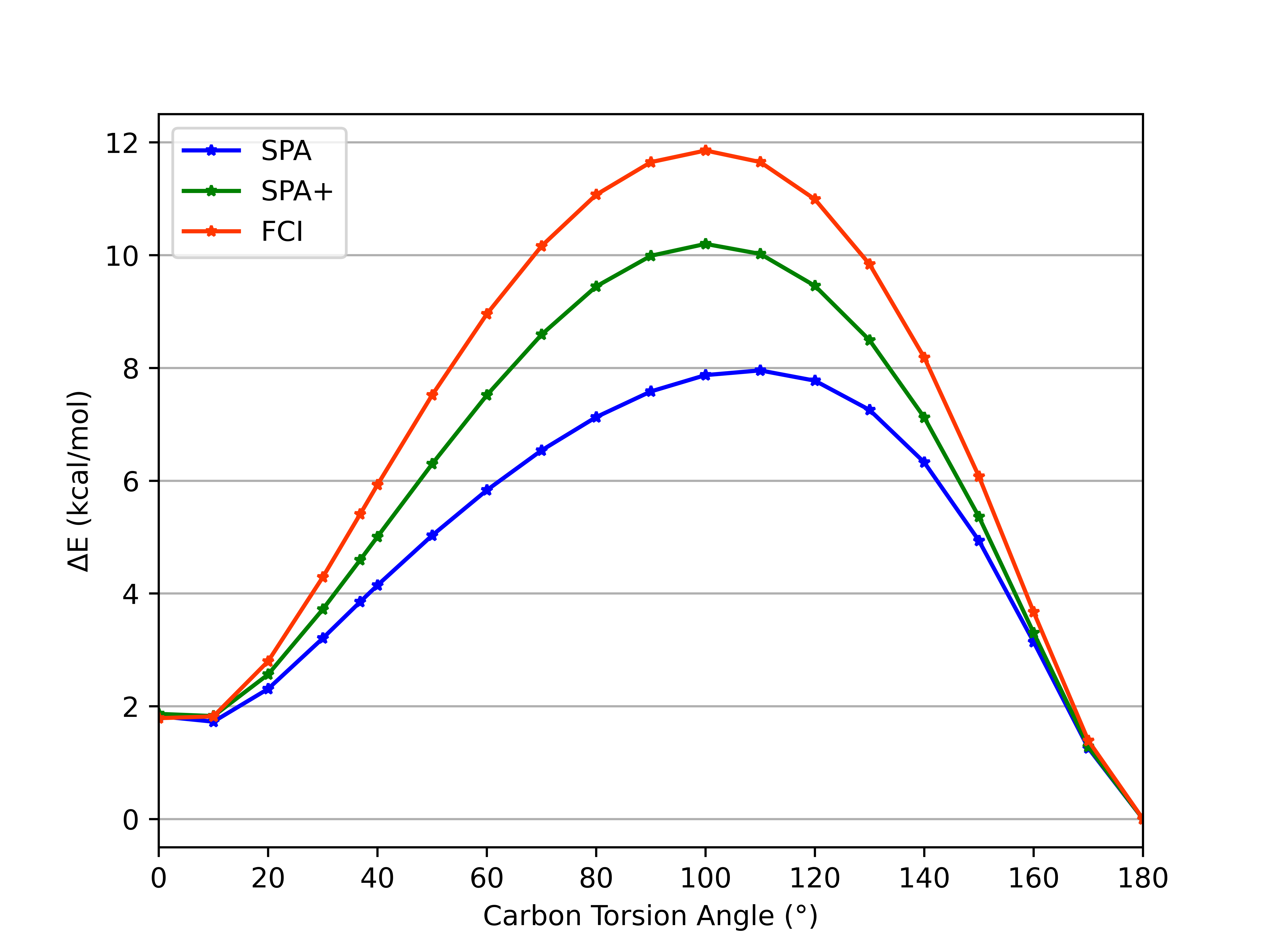}} \hfill
    \subfloat[\label{sfig:but_com}]{\includegraphics[width=0.49\linewidth]{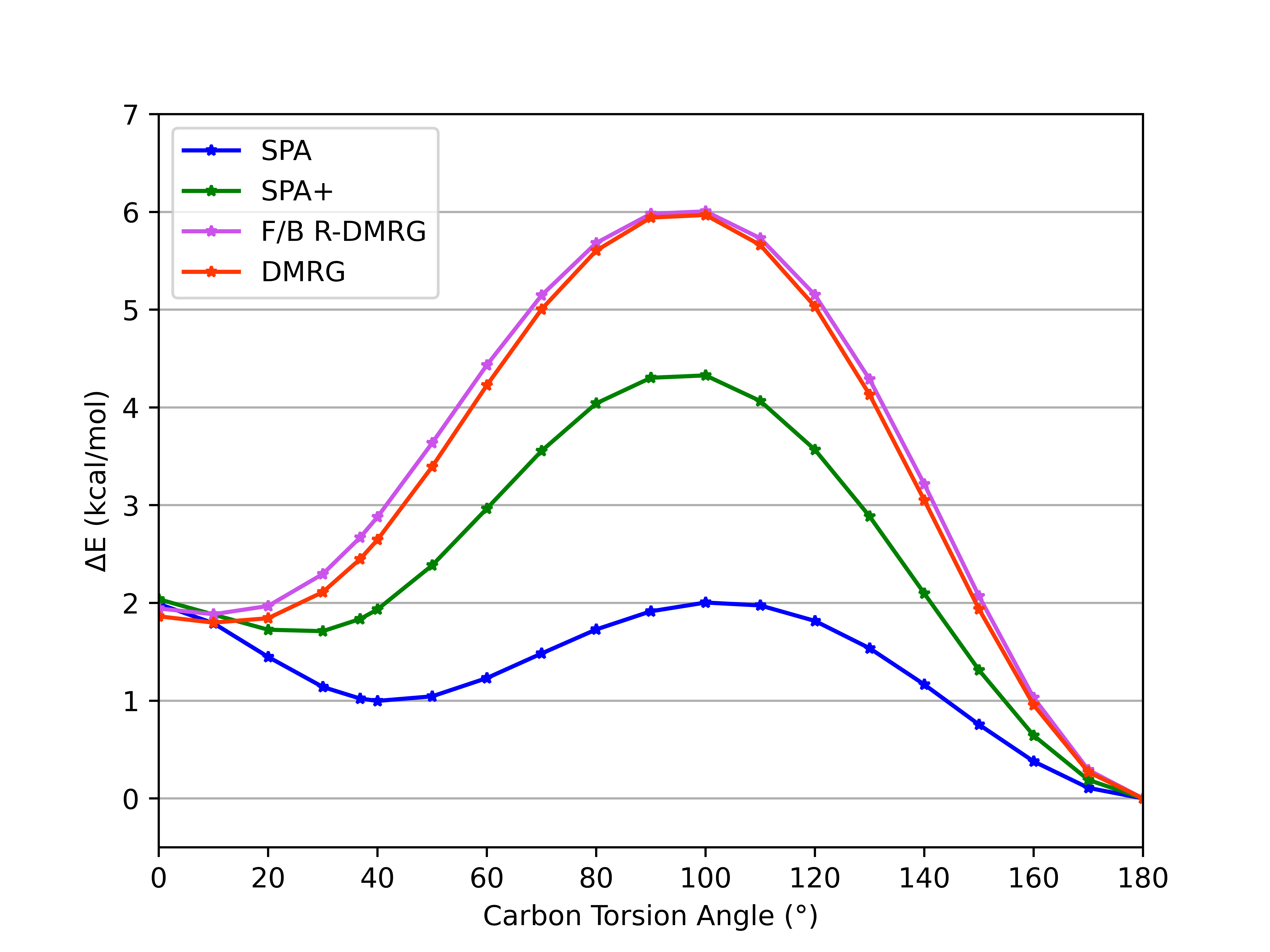}} \\
    \subfloat[\label{sfig:but_sep}]{\includegraphics[width=0.49\linewidth]{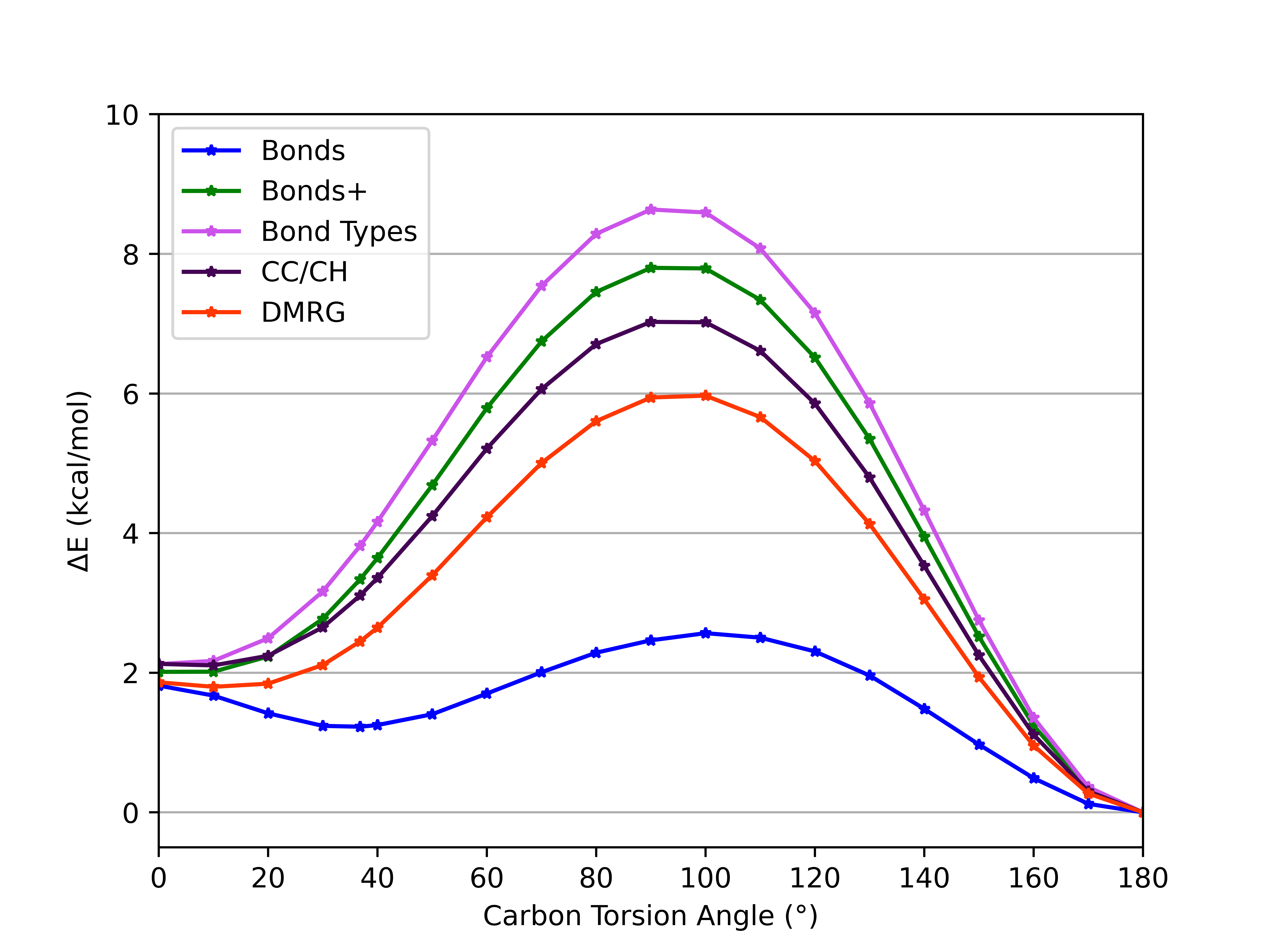}}
    \caption{Butadiene sto-3g carbon dihedral rotational energies, while keeping all other atoms' relative positions static. Energy difference with respect to trans (180°). All circuits are built as explained on Section~\ref{sec:VB_embedings}. \ref{sfig:but_act} Active space considering only the $\pi$-system in fermionic representation while all other electrons are frozen at Hatree-Fock level. \ref{sfig:but_com} Full molecule with $\pi$-system in fermionic and the rest in bosonic encoding. ``F/B R-DMRG'' correspond to the DMRG ground state calculation of the hybrid fermionic-bosonic Hamiltonian as defined on~\ref{eq:Hybrid}. In Fig.~\ref{sfig:but_sep}, DMRG energies on modified Hamiltonians as in Eq.~\ref{eq:Hsep} with the residuals dropped. In ``Bonds'' all bonds are decoupled (comparable to full SPA circuits), ``Bonds+'' treats the $\pi$-system collectively (comparable to SPA+). ``Bond Types'' the Hamiltonian is split into  C-C$\pi$, C-C$\sigma$ and C-H$\sigma$ orbitals and all couplings between the groups where removed. ``CC/CH'' treats the C-C$\pi$ and C-C$\sigma$ orbitals as one coupled group decoupled from the C-H$\sigma$ orbitals. ``DMRG'' denotes the complete fermionic Hamiltonian calculation.} 
    \label{fig:but_rotation}
\end{figure*}
In the current section, some application examples of the presented protocol will be presented. Unless otherwise stated, the classical reference will always be the Density Matrix Renormalization Group (DMRG).~\cite{schollwoeck2011Densitymatrix,chan2011Density,baiardi2020Density,yanai2015Density}
Here we assume that the use of localized orbitals that we obtain from the orbital optimizer will lead to good results.\cite{mitrushchenkov2012Importancea}  Moreover, this method naturally introduces the use of custom operators through the construction of Matrix Product Operators (MPOs), which is particularly useful in this work to study the errors associated to the hybrid encoding and wave function separability.
\subsection{Butadiene}\label{subsec:butadiene}
We chose butadiene as a demonstrative example of the method, given that although it is still a relatively small system for most classical quantum chemistry methods, it is already challenging for most quantum computing approaches. At the same time, it is one of the smallest potential examples for the valence bond embeddings introduced here. With the cis-to-trans rotational barrier this molecule also provides an initial example for a (intrinsic) relative energy. On Fig.~\ref{fig:but_rotation}, two main series are presented. In Fig.~\ref{sfig:but_act} we describe the results, when only the $\pi$-system (4 electrons in 4 spatial orbitals) is part of the active space with all other electrons frozen at Hartree-Fock level, while in Fig.~\ref{sfig:but_com} we describe the same approach with the $\sigma$-framework represented as hardcore-bosonic SPA wavefunction. In both cases, the $\pi$ system is represented by the same circuits, one a simple SPA and the other a multi-graph approach that takes a further resonance structure into account. Those circuits are identical to the circuits on linear $H_4$ in Ref.~\cite{kottmann2022molecular} As a third state, the $\pi$-system is represented either represented exactly (FCI on the active space approach) or, in case of the valence bond embedding with a DMRG on the hybrid Hamiltonian (F/B R-DMRG). The latter is used as a stand-in for the best possible result on used the hybrid encoding. 

In the active space approach we observe a faulty convergence in the transition energy, where the energy error with respect to the DMRG reference increases with more accurate descriptions of the $\pi$-system. In the valence bond embedding the contrary can be observed. 

The F/B R-DMRG is close to the unrestricted DMRG result, verifying the validity of the chosen separation into $\sigma$ (Bosonic) and $\pi$ (Fermionic) frameworks and the shortcomings in the chosen VQE circuit (SPA+). 
For this purpose we included some deeper analysis of the origin of this error. In Figure~\ref{sfig:but_sep},  DMRG energy for custom Hamiltonians has been computed, following equation~\ref{eq:Hsep}. On the one hand, blue and green series, labelled as \textit{Bond} and \textit{Bond+} respectively, correspond to the system separability defined by the \textit{SPA} and \textit{SPA+}, which fundamentally relies on the molecule bond structure, and taking into account the delocalization for the second case. Expected behaviour is found here, on the \textit{Bond} results, since it is incapable of incorporating the second graph contribution, seriously underestimating the rotational barrier. This would justify the considerable difference by just entangling the $\pi$ system. However, by only coupling these bonds, it significantly overestimates the barrier, since the electron density is unable to delocalize around the molecule. This can be shown on the tendency from the curve \textit{Bond Types} to \textit{CC/CH} and finally to the full \textit{DMRG}. In the serie \textit{Bond Types}, the system is split in C-H $\sigma$, C-C $\sigma$ and C-C $\pi$ orbitals; meanwhile, on the \textit{CC/CH}, the C-C $\sigma$ and C-C $\pi$ orbitals are entangled together. 
\subsubsection{Increasing the accuracy of the basis}
\label{sssec:bsc}
\begin{table}[!ht]
    \centering
    \caption{Butadiene energy in \textit{kcal/mol} difference w.r.t 180° for basis sets and energy methods. MRA refined bases have the same size as STO-3G}
    \begin{tabular}{lcc}
    \toprule
        \textbf{Angles (°)} & \textbf{0} & \textbf{90} \\
    \midrule
        \textbf{CCSD(T)}/sto-3g & 1.860 & 5.991 \\ 
        \textbf{CCSD(T)}/cc-pVDZ & 3.565 & 5.814 \\ 
        \textbf{CCSD(T)}/cc-pVTZ & 3.444 & 5.740 \\ 
        \textbf{SPA}/sto-3g & 1.990 & 1.915 \\ 
        \textbf{SPA+}/sto-3g & 2.038 & 4.304 \\ 
        \textbf{DMRG}/sto-3g & 1.863 & 5.943 \\ 
        \textbf{SPA}/MRA & 3.905 & 3.271 \\ 
        \textbf{SPA+}/MRA & 3.856 & 4.642 \\ 
        \textbf{DMRG}/MRA & 3.850 & 5.824 \\
    \bottomrule
    \end{tabular}
    \label{tab:but}
\end{table}
For most of the real-world relevant applications, a minimal basis such as STO-3G doesn't provide accurate results. A better spatial description of the wave function is required, usually addressed by increasing the basis set. In quantum computational methods, this will lead to increased qubit requirements as well as resulting increases in circuit depths. For this approach, the SPA parts are probably negligible as the corresponding circuits and parameters will increase linear, but potential algorithms downstream (such as phase estimation) will be impacted more severly. In order to mitigate this issue, we have chosen to improve the basis quality rather than its size. This can for example be achieved through orbital refinement in a multiresolution real-space grid.~\cite{langkabel2024adventfullyvariationalquantum,frayedends} \\\\
In order to study how this predicted property is affected by the basis limitation, Table~\ref{tab:but} presents the energy difference (in \textit{kcal/mol}) computed for butadiene with respect to the trans structure for a series of energy methods at different bases. On the classical computing side, we chose \textit{CCSD(T)} as it is cheaper than DMRG and performs quite well for organic molecules: For the minimal basis, the CCSD(T) results agree with \textit{DMRG}. The main result presented here is that, even if our approach is still around one \textit{kcal/mol} from the reference, it behaves consistently with the basis set refinement. 
\subsection{Benzene and Naphthalene}
\begin{figure*}
    \centering
    \subfloat[]{\label{sfig:benzal_str}\includegraphics[width=0.8\linewidth]{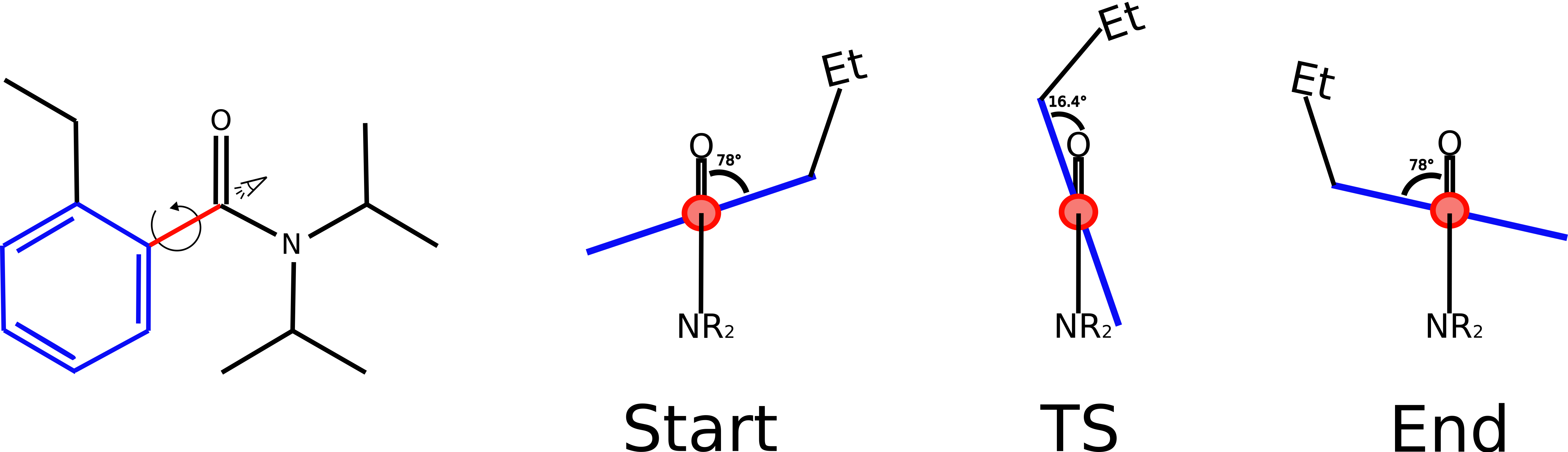}}
    \\ 
    \subfloat[]{\label{sfig:benzal_res}\includegraphics[width=0.8\linewidth]{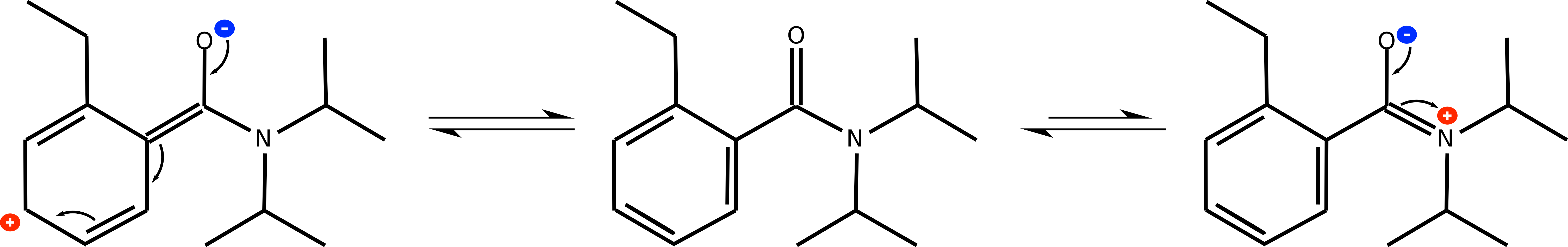}}
    \caption{\ref{sfig:benzal_str} 2-Ethyl-N,N-diisopropylbenzamide geometrical structures. \ref{sfig:benzal_res} Resonant structures (graphs) considered in the SPA+.}
\end{figure*}
\begin{table}[!ht]
    \centering
    \caption{SPA+ circuit compilation results, reporting counts of single-qubit gates and \textit{CNOT}s, together with the number of fully commuting groups for both the complete hybrid Hamiltonian (Eq.~\ref{eq:Hybrid}) and its separable counterpart (Eq.~\ref{eq:Hsep}). Commuting groups are obtained via the \textit{Sorted Insertion} method~\cite{crawford2021Efficienta,bansingh2022Fidelity} under the Qubit-Wise Commuting criterion~\cite{verteletskyi2020measurement}.}
    \begin{tabular}{cccc}
    \toprule
        \textbf{} & \textbf{Butadiene} & \textbf{Benzene} & \textbf{Naftalene} \\ 
        \textbf{Depth} & 184 & 214 & 428 \\ 
        \textbf{CNOTs} & 203 & 319 & 648 \\ 
        \textbf{Single Qubit Gates} & 262 & 390 & 648 \\ \midrule
        \textbf{$\hat{\mathbf{H}}$ Groups} & 114 & 210 & 2557 \\ 
        \textbf{R-$\hat{\mathbf{H}}$ Groups} & 27 & 85 & 88 \\ 
        \bottomrule
    \end{tabular}
    \label{tab:comp}
\end{table}
To assess how the separable approach scales to larger molecular systems, benzene and naphthalene were selected as initial test cases due to their resonance structures, which are analogous to those discussed previously. However, the increased size of the fermionic subsystem (half-filling models of 12 and 20 qubits) already introduces significant computational overhead. To illustrate this, Table~\ref{tab:comp} reports circuit specifications and the number of commuting groups. The former is obtained through compilation into single-qubit gates and \texttt{cnot} operations (following the compilation strategies of~\cite{kottmann2022Optimized,yordanov2020efficient}), and the number of fully commuting groups are obtained via the Sorted Insertion method~\cite{bansingh2022Fidelity} under the Qubit-Wise Commuting heuristic~\cite{verteletskyi2020measurement}.\\\\
It is important to note that the Hamiltonians considered here are not the full fermionic Hamiltonians, but rather the hybrid fermionic/bosonic form in eq.~\ref{eq:Hybrid}. These were grouped both without separability constraints ($\hat{H}$ terms) and with separability enforced (R-$\hat{H}$ terms). The results indicate that even this minimal setup leads to substantial circuit depth and a large number of \textit{CNOT} gates; the circuit depths are however significantly reduced when compared to regular VQE approaches (see~\cite{kottmann2023Molecular}). Additionally, the separability constraint significantly reduces the number of commuting groups, highlighting the practical advantage gained by discarding non-separable terms. 
\begin{table}[!ht]
    \centering
    \caption{Energy Errors in ‰ with respect to DMRG on the full Fermionic Hamiltonian for a sequence of VQE and restricted-DMRG energies.}
    \begin{tabular}{lccc}
    \toprule
        \textbf{} & \textbf{Butadiene} & \textbf{Benzene} & \textbf{Naftalene} \\ 
        \textbf{SPA} & 0.787 & 2.481 & 2.459 \\ 
        \textbf{SPA+} & 0.762 & 2.252 & 2.298 \\ 
        \textbf{DMRG/(Bonds)} & 0.295 & 1.713 & 1.745 \\ 
        \textbf{DMRG (Bonds+)} & 0.239 & 0.787 & 0.822 \\ 
        \textbf{DMRG (F/B)} & 0.714 & 1.333 & 1.388 \\ 
        \textbf{DMRG (Bond Types)} & 0.169 & 0.576 & 0.506 \\ 
        \textbf{DMRG (CC-CH)} & 0.156 & 0.558 & 0.485 \\ \bottomrule
    \end{tabular}
    \label{tab:sep}
\end{table}
However, it would still need to address whether the separability constraint is suitable for these more complex systems. For this purpose, we have computed DMRG energies with the custom operators as presented on Fig.~\ref{sfig:but_sep}. Relative error energies (in ‰) with respect to the complete DMRG have been presented on Table~\ref{tab:sep}, computed as
\begin{equation}\label{eq:relerror}
    \epsilon_i = \frac{E_i-E_{DMRG}}{E_{DMRG}}*1000.
\end{equation}

The analysis verifies the original approximation of Fig.~\ref{fig:sep_circuit} (Bonds+ series, that represents the $\sigma$/$\pi$ splitting without further approximations), but and reveals that the hardcore-bosonic approximation is more suitable for the C-H parts of the molecule. 

\subsection{Usecase}
\begin{table}[!ht]
    \centering
    \caption{Benzamide rotation barrier in kcal/mol. SPA contains only the main graph (central structure on~\ref{sfig:benzal_res}). SPA+1 adds the graph with the positive charge on the Nitrogen (right). SPA+2 adds positive charge delocalization on the benzene ring (left). SPA+12 combines all three structures.}
    \begin{tabular}{lc}
    \toprule
        \textbf{} & \textbf{Rotational Barrier (kcal/mol)} \\ \midrule
        \textbf{CCSD(T)} & 13.5 \\ 
        \textbf{SPA} & 14.3 \\ 
        \textbf{SPA+G1} & 14.3 \\ 
        \textbf{SPA+G2} & 13.4 \\ 
        \textbf{SPA+G1+G2} & 13.4 \\ 
        \textbf{DMRG} & 13.1 \\ \bottomrule
    \end{tabular}
    \label{tab:benz_rot}
\end{table}
To further probe the limits of the proposed framework, the rotational barrier of 2-Ethyl-N,N-diisopropylbenzamide ($C_{15}H_{23}NO$) was simulated, as shown in Fig.~\ref{sfig:benzal_str}. This molecule is not only larger and contains heteroatoms, but is also of particular interest because three resonance structures are expected to contribute. Due to steric hindrance this also represents a $\pi$-system with limited conjugation: due to the non-planar equilibrium conformers the impact of the resonance structure on the left Fig.~\ref{sfig:benzal_str} (conjugation of the peptide group with the ring). Table~\ref{tab:benz_rot} reports the rotational barrier computed with various methods. Note, that we did not distinguish between forward and backward barrier, as \textit{Start} or \textit{End} conformers are close in energy (below the millihartree threshold). The results indicate that, for this property, the resonance structure featuring a positively charged nitrogen on the peptide group contributes only marginally and the conjugation with the ring has significant impact on the transition state -- witnessed through the overestimation of the rotational barrier when the resonance structure is absent. In summary this examples demonstrates that our approach provide chemically interpretable methods with reasonable level of accuracy, while maintaining a computational cost that remains within practical limits.  

\section{Conclusion \& Outlook}
In this work, we integrate concepts from Valence Bond theory with a hybrid encoding scheme to increase the applicability of variational quantum eigensolvers. In particular we have significantly lowered the computational cost with a valence bond embedding that allows classical simulation of large fractions of the total circuit as well as the range to which such methods can be applied to, demonstrating this on some of the largest VQE instances up to date. By building on top of the separable pair approximation we include a baseline level of accuracy that is classically simulable (when all clusters are at SPA level) suitable for subsequent methodological refinements. Through the combination of localized orbitals with the Quantum Valence Bond framework we have developed a particularly effective framework, as it enables a natural and problem-adapted distribution of computational resources.

Moreover, we present numerical simulations on molecular systems that exceed the sizes typically accessible within standard approaches, while still leaving room for further methodological improvements. Finally, we illustrate that the proposed framework extends naturally toward the basis-set limit through the use of multiresolution analysis bases, without requiring substantial modifications to the underlying ansatz. We incorporated our algorithms within the open-source package \textsc{project-sunrise}~\cite{sunrise} where they can hopefully be leveraged within future methodologies.

\section*{Scientific Software}
Development and data generation within this work have been conducted through the open-source package \textsc{tequila}~\cite{tequila} using \textsc{qulacs}~\cite{qulacs} as simulation backend, the JW transformation from \textsc{open-fermion}~\cite{OpenFermion}, molecular integrals as well as most of classical methodology from \textsc{pyscf}~\cite{pyscf1,pyscf2, pyscf3}, DMRG calculations from \textsc{block2}~\cite{zhai2023Block2} and the automatically differentiable framework described in Ref.~\cite{kottmann2021feasible}. Calculations with MRA representation used \textsc{madness}~\cite{harrison2016madness} via \textsc{frayedends}~\cite{frayedends} following the descriptions in Ref.~\cite{kottmann2020direct, kottmann2020reducing, langkabel2024adventfullyvariationalquantum}. \textsc{clpo} orbitals were generated using \textsc{janpa}~\cite{nikolaienko2014JANPA,nikolaienko2019Localizeda}. Quantum circuits are created via \textsc{qpic}~\cite{qpic}.\\\\
All the hybrid encoding~\cite{delarcosantos2025Hybrid}, \textsc{janpa} interface, as well as all the separable circuit compilation are available on \textsc{project-sunrise}.~\cite{sunrise}

\section*{Acknowledgement}
The authors acknowledges support from the Federal Ministry of Research, Technology and Space (BMFTR) of Germany through the VeriVaQ project.
JSK gratefully acknowledges support from the Hightech Agenda Bayern and the Munich Quantum Valley. We thank Davide Bincoletto for various discussions. Computational resources were in-part provided by the LiCCA HPC cluster of the University of Augsburg, co-funded by the German Research Foundation (DFG) – Project-IDs 499211671 \& 572310035.
\bibliography{main.bib}
\end{document}